\documentclass[11pt,twoside,A4]{article}
\usepackage{times,fancyhdr}
\usepackage[dvips]{graphicx}
\usepackage{latexsym}
\usepackage[affil-it]{authblk}
\usepackage{mathptm}
\usepackage{amsmath}
\usepackage{amssymb}
\usepackage{floatrow}
\usepackage{units}
\usepackage{setspace}
\usepackage{hyperref}
\usepackage[font=small,labelfont=bf]{caption}
\usepackage[top=4cm, bottom=4cm, left=3.5cm, right=3.5cm]{geometry}

\def\beq{\begin{equation}}
\def\eeq{\end{equation}}
\def\beqn{\begin{eqnarray}}
\def\eeqn{\end{eqnarray}}

\begin{document}

\title{Superdeterminism: A Guide for the Perplexed}
\author{Sabine Hossenfelder}
\affil{Frankfurt Institute for Advanced Studies\\
Ruth-Moufang-Str. 1,
D-60438 Frankfurt am Main, Germany
}
\date{}
\maketitle
\vspace*{-1cm}

\begin{abstract}
Superdeterminism is presently the only known consistent description of nature that is local, deterministic,
and can give rise to the observed correlations of quantum mechanics. 
I here want to explain what makes this approach promising and offer the reader some advice for
how to avoid common pitfalls. In particular, I explain why superdeterminism is not a threat to science,
is not
necessarily finetuned, what the relevance of future input is, and what the open
problems are. 

\end{abstract}

\section{Introduction} 

From their student days on, physicists are used to hearing that one cannot understand quantum mechanics,
that quantum mechanics is strange, weird, counterintuitive, and that some questions should not be asked. 
I here want to explain how the strangeness disappears if one is willing to accept that one of the 
assumptions we have made about quantum mechanics is not realized in nature: Statistical Independence. 

Theories that do not
fulfill the assumption of Statistical Independence are called ``superdeterministic'' and they are
surrounded by a great deal of confusion. One of the reasons is that historically the debate around superdeterminism has
been mixed up with the question whether we have free will. Indeed,
the assumption of Statistical Independence is frequently referred to as the ``free will'' or ``free choice''
assumption. The
question of free will, however, has nothing to do with the ``super'' in superdeterminism,
but is an issue that generally comes up in deterministic descriptions of nature. This point
was already addressed in \cite{Hossenfelder:2019shy}.

I here want to leave aside the free-will debate because for what the physics is concerned I frankly don't think it's particularly interesting. Instead, I want to give some practical advice to the reader who is interested in understanding, or maybe
even working on, superdeterminism. 

\section{How to think about superdeterminism without tying your brain in knots}
\label{knots}

The defining properties of superdeterministic models are that they are Psi-epistemic, deterministic, local, hidden variables models which
violate Statistical Independence. For a detailed definition of these terms, see \cite{Hossenfelder:2019shy}. A superdeterministic
model has a deeper, underlying level, from which quantum mechanics derives on the average. Superdeterminism, therefore,
is not an interpretation of quantum mechanics. It is a more fundamental description of nature, and Psi-epistemic not
because it is non-realist, but because the wave-function (``Psi'') of quantum mechanics is an average description, hence epistemic and not ontic.

The one unfamiliar property of superdeterminism is the violation of Statistical Independence. Loosely speaking, Statistical
Independence means that the degrees of freedom of spatially separated systems can be considered uncorrelated,
so in a superdeterministic model they are generically correlated, even in absence of a common past cause. 
In the context of Bell's theorem \cite{Bell} in particular, one
has a correlation between the hidden variables, which determine the measurement outcome, and the detector settings. 

Statistical Independence is an assumption that we make about nature which works well to describe our observations on the
macroscopic ``classical'' level. We do not know, however, whether Statistical Independence is fundamentally correct,
and superdeterminism is based on the premise that it is not. 
Since it is only with this assumption that the observed violations of Bell's inequality force us to give up a local,
deterministic completion of quantum mechanics, we should find out what happens if we drop it. That is because
once we
drop Statistical Independence, there is nothing preventing us from developing a consistent local and deterministic theory
underlying quantum mechanics, and this is clearly something that we should try before accepting that nature
is fundamentally unpredictable and ununderstandable, which amounts to giving up. 

In mathematical terms, Statistical Independence can be expressed as a correlation between the hidden
variables of the theory, usually denoted $\lambda$, and the detector settings of an experiment, $\theta$. 
Both $\lambda$ and $\theta$ could contain several degrees of freedom, so the reader should understand
them as collective terms. In particular, $\theta$ might describe the settings of multiple, spacially separated, detectors,
as we commonly have in {\sc EPR}-type experiments.

The hidden variables, in Bell's notation, are all the information that is required to predict the outcome
of a measurement except for the measurement settings. If one does not know the hidden variables, 
then one can merely make a probabilistic prediction based on the measurement settings. 
This probabilistic description is the one we normally get in quantum mechanics.

It must be emphasized that $\lambda$ is not necessarily a property of the prepared state. It is not
even necessarily localized in any meaningful sense. It is really any kind of information that allows one to make a prediction.
The term ``hidden'' merely means that these variables do not explicitly appear in standard 
quantum mechanics. There is no reason why these  variables should be per se unmeasurable. In fact, since they determine the measurement outcome, one can use the measurement
outcome to make inferences about the hidden variables, at least if they are computable. We will come
back to this point later.

One may ask if not $\lambda$ could be time-dependent, $\lambda(t)$. Yes, it could be time-dependent, but taking this
possibility into account does not constitute a new case.\footnote{As already pointed out by Bell \cite{Bell:1987hh}.} That is because the theory is deterministic, 
so there must be an evolution law that allows one to compute $\lambda(t)$ from an initial value $\lambda_0$. 
Therefore, we could replace $\lambda(t)$ with $\lambda_0$ and then drop the index. In other words, considering
$\lambda$ to be a collection of constants is already the most general case, provided the theory is deterministic.\footnote{I should mention that some 
authors do not require a superdeterministic theory to also be deterministic. This ignores the original reason to consider 
superdeterminism and defeats the point of thinking about it. Therefore, while it is an option that mathematically exists, it is
not one I consider here.} 

However, labeling a time-dependent evolution of a state by its initial value is rarely of
practical use for making predictions. If you throw a ball and want to know where it will come down, then saying it's determined by the initial
position and velocity is correct, but not very insightful. For the same reason, while the case with time-dependent $\lambda$ is technically subsumed in the case with constant $\lambda$, expressing a theory in term of constant $\lambda$ may sometimes not be practically useful. 

The way that Statistical Independence makes its appearance in superdeterminism is that the probability distribution of the hidden variables given the
detector settings $\rho(\lambda |\theta)$ is not independent of the detector settings, ie $\rho(\lambda | \theta) \neq \rho (\lambda)$.
 What this means is that
if an experimenter prepares a state for a measurement, then the outcome of the measurement will depend on the
detector settings. The easiest way to think of this is considering that both the detector settings, $\theta$, and the hidden variables, $\lambda$, enter the evolution law of the prepared state. As a consequence, $\theta$ and $\lambda$ will generally be correlated at the time of measurement, even if they were uncorrelated at the time of preparation. 
Superdeterminism, then, means that the measurement settings are part of what determines the outcome of the time-evolution of the prepared state.

What does it mean to violate Statistical Independence? It means that fundamentally everything in the
universe is connected with everything else, if subtly so. You may be tempted to ask where these connections come from,
but the whole point of superdeterminism is that this is just how nature is. It's one of the fundamental assumptions
of the theory, or rather, you could say one drops the usual assumption that such connections are absent. 
The question for scientists to address is not why nature might choose to violate Statistical Independence, but
merely whether the hypothesis that it is violated helps us to better describe observations.

I believe that many people find violations of Statistical Independence confusing because such violations have frequently been
portrayed in the literature in an extremely misleading way. Superdeterminism is supposedly a ``conspiracy'' in which the detector
settings are influenced by hidden variables in just the right way to give the results of quantum mechanics. But that 
Statistical Independence is violated tells us nothing about the freedom to choose measurement settings, unless you
want to say that the measurement setting tells you what the measurement setting is, which is correct, but not very deep.
In a superdeterministic theory, the measurement setting, $\theta$, can be chosen in the same way you can chose it in normal quantum mechanics. It's
just that what the prepared state will evolve into a depends on those measurement settings. 

Part of this confusion comes from mistaking a correlation for a causation. Violations of Statistical Independence,
$\rho(\lambda | \theta) \neq \rho (\lambda)$, do not tell us that $\lambda$ influences $\theta$. They merely tell us
that $\lambda$ and $\theta$ are correlated.

Another part of this confusion stems from forgetting that Bell's theorem makes statements about the {\emph{outcomes}} of
measurements. It does not have a time-dependence. So really all the variables which appear in the theorem are
the variables at the time of measurement. This also immediately clears up the confusion about what happens
if one changes the detector settings between preparation and detection. The answer is: nothing. Because the
setting at preparation was always irrelevant, and so was the exact way in which these settings came about. 
In a superdeterministic theory, the evolution of the prepared state
depends on what the detector setting is {\emph{at the time of measurement}}. 

I want to stress that this is not an interpretation
on my behalf, this is simply what the mathematics behind Bell's inequality really says. $\rho(\lambda | \theta) \neq \rho (\lambda)$,
put into words, means that the hidden variables which determine the measurement outcome are correlated with the detector settings at the time of measurement. 
This does not necessarily mean the hidden variables were correlated with the measurement settings at the time of preparation. This correlation may instead have
come about dynamically.

However, if the theory is deterministic, one can replace the state of the detector at the time of
measurement with the state of the detector at the time of preparation plus the evolution law. This makes the
situation dramatically more complicated because to find out what the setting will be at the time of 
detection one now has to calculate what goes on
in the experimenter's brain. If one in this way insists
on describing the measurement setting at the time of detection by the state of the system at an earlier
time plus a hideously complicated evolution of a macroscopic system, then the violation of Statistical Independence begins 
to appear conspiratorial. It seems that, somehow, something must have made the experimenter change the settings 
so that they ``fit'' to the evolution of the prepared state, when really the only thing that matters
was the setting at the time of measurement all along. 

One can make the situation even more complicated by rolling back the initial 
condition further in time by some billions of years so that now one has to take into account each and
everything that happened in the backward lightcone of the actual measurement. It then seems like
one needs to arrange the hidden variables in ``just the right way'' so that any possible method for choosing
the detector settings comes out correctly. This is one of the reasons that superdetermism
is widely believed to be finetuned. 

I will explain in Section \ref{finetuning} why this and other finetuning arguments are wrong, but first let me
say some more about what it means to violate Statistical Independence. 

\section{The role of non-linearity for superdeterminism}

In quantum mechanics, the dynamical law (ie, the Schrödinger equation) together with Born's rule, does not predict the outcome of a
measurement; it merely gives a probabilistic prediction. All the trouble with interpreting quantum mechanics
comes from taking this probabilistic prediction to be fundamental, rather than a consequence of missing
information. 

This is because if one takes the probabilistic prediction to be fundamental (ie takes a Psi-ontic perspective) then one has to accept that the
probabilistic prediction cannot be an ensemble prediction because it must be meaningful also for individual measurements on single particles. 
With this comes the problem that once we have detected a particle in a particular detector eigenstate, the probability that it is in the measured state equals 1, which is generically not
what quantum mechanics predicts (unless the state happened to have been prepared in this state already). Therefore, the wave-function which describes the system must be updated upon measurement 
so that it agrees with observation. This measurement
update is also sometimes called the ``collapse'' of the wave-function. 

Decoherence does not remove the need to update the wave-function. Decoherence merely converts a coherent quantum state into
a classical probability distribution. But to describe what we observe, the state would have
to evolve into the actually measured outcome with probability (almost) 1. Attaching the measurement postulate to the Schrödinger equation gets the job
done. However, this procedure is internally inconsistent. If quantum mechanics was fundamental,
then what happens during measurement should be derivable from the behavior of the particles that the detector
is made of. It should not be necessary to have two separate laws. The measurement
postulate, however, cannot follow from the Schrödinger equation because it is non-linear. This is known
as the ``measurement problem'' of quantum mechanics.

There are only two ways to resolve this inconsistency. The one advocated here is that
the measurement postulate is itself an effective description of a process in an underlying theory. That is,
quantum mechanics is not fundamental, but derived from a deeper natural law, and the indeterminism
of quantum mechanics is due to lack of knowledge of details of the system. 

The other logically possible way to solve the
measurement problem is to give up on reductionism and postulate that the behavior of a measurement 
device cannot be derived from the behavior of its constituents. This would require developing a theory
that quantifies when and how reductionism breaks down. I am not aware that such a theory exists and
it is highly questionable that if it existed it would be compatible with observations. For better or worse, if
we take apart detectors, they are ostensibly made of particles. It is certainly possible to posit that the
behavior of a system changes with the number of constituents, but note that in this case the laws of
the composite system still derive from the laws for the constituents, just not in the way that we currently
expect. Indeed, it is difficult to even make mathematical sense of what it means for reductionism to break down.

All interpretations of quantum mechanics that take it that the wave-function (or density matrix, in the
more general case) encodes knowledge held by observers suffer from the same problem. They are
either postulating a process (the update) that should be derivable from the theory's dynamical law,
or they are implicitly assuming that reductionism breaks down, yet not telling us when or how. 

Let us therefore look at the scientifically more interesting option that quantum mechanics is not fundamental and the ``collapse''
of the wave-function is merely an effective description of an underlying process. A most important
clue to understanding how this underlying theory looks like is to note that the measurement process is non-linear. I can
prepare a system in state $|\Psi_A \rangle$ so that the outcome $A$ will be measured with probability 1. 
I can also
prepare a system in state $|\Psi_B \rangle$ so that the outcome $B$ will be measured with probability 1. 
But if I prepare the state ${ \left(|\Psi_A \rangle + |\Psi_B \rangle \right)}/{\sqrt{2}} $, the measurement outcome
will {\emph{not}} be a superposition of the outcomes of the individual prepared states. Instead, it will
be either $A$ or $B$ with probability 1. 

So whatever the dynamical law of the underlying theory is, it has to be non-linear in the ontic states of
the underlying theory, whereas suitably averaged it gives rise to the linear evolution of quantum mechanics
plus the sudden measurement update. That is to say, we do not want to make the Schrödinger equation
non-linear, we are looking for an underlying non-linear theory that gives rise to the ordinarily linear evolution law
of quantum mechanics. The wave-function itself is already an ensemble-description. It is not the ontic state
for which we seek a non-linear evolution law.

As also discussed in \cite{Hossenfelder:2019shy}, this approach fits well with the observation that the Schrödinger 
equation can be equivalently
rewritten into the von Neumann-Dirac equation which looks extremely similar to the Liouville
equation. The Liouville equation, however, is an equation for statistical ensembles and linear in the
phase-space density regardless of whether the underlying theory is linear.\footnote{This similarly between
quantum mechanics and statistical mechanics has been made formally precise in a variety ways, 
see e.g. \cite{Moyal,Wigner1,Wigner2,Gibbons}, though these approaches run into difficulties with the
probabilistic interpretation. The origin of this difficulty is almost certainly that these so-far conceived statistical
underpinnings of quantum mechanics are not superdeterministic.}

In the superdeterministic evolution, the detector
eigenstates have to be outcomes of the time-evolution with a probability of almost 1 just because that is the
only thing we ever measure. The theory therefore has
a non-linear time-evolution of which the detector eigenstates are attractors. I here use the word
``attractor'' to mean that the time-evolution of {\emph{any}} initial state will tend to a 
state nearby a subset of all possible states. This subset generically consists of several disconnected
components which correspond to the detector eigenstates. 

Superdeterminism enters this time-evolution by requiring that {\bf{the locations of the attractors in configuration-space
depend on the measurement settings}}. So we arrive at a picture where instead of having one
prepared state, we have an ensemble (epistemic) of prepared states (ontic) with a probability distribution over hidden variables. The
 system is only in one of these states, but we do not know in which one. Each
state in the ensemble will evolve into a detector eigenstate with probability 1, regardless of what the exact value of the 
hidden variable.
But the hidden variable itself determines which detector eigenstate is the outcome. 

To recap: If you know the exact value of the hidden variable, then the state will evolve into
a particular detector eigenstate with probability 1. Different detector settings correspond to different
 attractor locations. If you do not know the exact value of the hidden variables, then you
can only make a probabilistic prediction.  For an example of such
an attractor dynamics for a simple case, see \cite{Palmer:1995mxd,toymodel}.

It should be mentioned here that one can make superdeterminism non-linear by
a superselection rule rather than by a dynamical law. This basically means that one assumes a particular
basis of states is ontic and superpositions of this basis do not occur in nature. This assumption itself
is non-linear and used in the celluar automaton approach proposed by `t Hooft \cite{Hooft:2014kka}. 

\section{Superdeterminism is neither classical, nor realist, nor an interpretation of quantum mechanics}

Superdeterministic theories are not classical in any meaningful sense of the 
word. They can, and frequently do, employ the mathematical apparatus of quantum theory, as with states of the system being
vectors in Hilbert spaces and observables being defined by hermitian operators acting on the states, some
of which are non-commuting. Importantly,
this means that superdeterministic theories can contain entangled states the same way that quantum
mechanics does, and they also reproduce the uncertainty principle for conjugated variables. There is no need to
change any of this because these features of quantum mechanics are unproblematic. The problematic part is 
the measurement update, which is what superdeterminism eliminates.

I have also repeatedly encountered physicists who either praise superdeterminism for being a realist
interpretation of quantum mechanics or who belittle it for the same reason. As someone who is
not a realist themselves, I am offended by both positions. Superdeterminism is an approach
to  scientific modeling.
We use it to describe observations. (Or at least we try to.) Whether you believe that there is
something called ``reality'' which truly is this way or another is irrelevant for the model's scientific success. 

Realism is a philosophical stance. There is nothing in science itself that can tell you whether
the mathematical structures in our models are real or whether they merely describe observations.
It also doesn't matter. You could be a solipsist and I would still argue that you would be
better off with a superdeterministic model than with quantum mechanics because in
contrast to quantum mechanics, superdeterminism is internally consistent. 

Another confusion is that superdeterminism can't be right because it cannot faithfully reproduce the
predictions of quantum mechanics in all circumstances. In \cite{Landsman}, for example, it is argued that to reproduce quantum mechanics 
requires true randomness, in which case the theory cannot be deterministic. 

Leaving aside that the exact argument 
hinges on the definition of randomness used in \cite{Landsman}, which some may quibble with, the conclusion is 
correct. Superdeterminism does of course not perfectly reproduce quantum mechanics, but only recovers it in the limit 
where one assumes the hidden variables are really random. The whole point of considering superdeterminism is 
that it is a more fundamental theory which in some cases does {\emph{not}} give the same predictions 
as quantum mechanics. Complaining that superdeterminism does not perfectly reproduce quantum mechanics is like
complaining that statistical mechanics does not perfectly reproduce thermodynamics because $N \to \infty$ is not
a limit that exists in reality. The claim is technically correct, but one doesn't learn anything from it.

Superdeterminism, therefore, is not an interpretation of quantum mechanics, but an actual modification.
It is ``beyond quantum mechanics'' the same way that models for particle dark matter are ``beyond the
standard model''. In that, superdeterminism is most
similar to models with spontaneous collapse. In contrast to collapse models, however, superdeterminism
does not recover quantum mechanics by re-introducing randomness into the dynamical law. 

\section{Superdeterminism isn't as weird as they say} 

A lot of people seem to think that
violating Statistical Independence is weird, but is no weirder than quantum mechanics already is because 
in quantum mechanics too the measurement outcome depends
on the detector settings. Yes, it does: You need the detector setting to update the wave-function. You
need it to define the projection operators that enter the collapse postulate.  

The reason this doesn't appear in the math as a violation of Statistical Independence is that in
quantum mechanics we accept that the measurement outcome is not determined, so there are no 
hidden variables to correlate with anything. The price we pay is that instead of having a dynamical
law that depends on the detector settings and {\emph{locally}} gives rise to a measurement eigenstate, in
quantum mechanics we have a law that does not depend on the detector settings, but then, when the measurement
actually happens, it suddenly becomes undeniable that this was actually the wrong dynamical law. So then we
correct for our mistake by what we call the update of the wave-function. As a consequence, this update is instantaneous and
appears non-local, but that is because it never described the dynamics of the ontic states to begin with.

One sees this too in Bohmian mechanics and models with spontaneous collapse, both of which require
information about the detector settings to produce dynamics that agree with observation.\footnote{Bohmian mechanics
is non-local in some sense but what's non-local about it is not the dynamics of the supposed particles.} It's just
that followers of these approaches like to insist that really the only thing we ever measure are positions, in which
case there's no need to choose what you measure, hence nothing to postulate. Of course this is also why no one 
who works on quantum field theory takes either Bohmian mechanics or collapse models seriously. 

Again, though, the dependence on the measurement settings doesn't formally register in the math as violations of Statistical Independence, in these two cases
because the detector setting is not in what is normally said to be the ``hidden variables'' of these models (in Bohmian mechanics, 
the positions of the other particles, in collapse models the steps in the
random walk). But note that then these hidden variables do not by themselves determine the measurement outcome. 
To determine the outcome one also needs to know the detector settings.

\section{But someone told me superdeterminism would kill  science}

The idea that violations of Statistical Independence would spell the death of science seems to go back
to a claim by Shimony, Horne, and Clauser in 1976 \cite{SC}. It has since been repeated in slight 
variations many times. For quotes, see eg \cite{Hossenfelder:2019shy,Chen:2020yoa}
and references therein (or better still, read the originals). 

In a nutshell, the argument
has it that scientists use the assumption of Statistical Independence to analyze outcomes of 
random trials. Think about randomly assigning mice to groups for studying whether tobacco causes cancer, 
to name a common example. Without Statistical Independence one would be allowed to postulate that observed correlations 
(between tobacco exposure and cancer incidence) just
happened to be encoded already in the initial state of the experimental setup, science would falter, and the world 
would fall apart into shredded pages of {\it{Epistemological Letters}}, or so I imagine.

The problem with this argument is that, crucially, it leaves aside {\emph{why}} we use the assumption of Statistical Independence and {\emph{why}}
we do not consider its violations viable scientific explanations for random trials. Let me fill you in: We use the assumption of Statistical Independence
because it works. It has proved to be extremely useful to explain our observations. Merely assuming that Statistical Independence
is violated in such experiments, on the other hand, explains nothing because it can fit any data. 

Here, I am using
the word ``explain'' in a very narrow, computational sense. A theory ``explains'' data if it requires less detail as input than
data it can fit (to some precision). One could quantify the explanatory power of a theory, but for our
purposes that's an overkill, so here and in the following I use ``explain'' to loosely mean ``fitting a lot of data with few
assumptions''. This may be somewhat vague but will do.

So, it is good scientific practice to assume Statistical Independence for macroscopic objects because it explains data and thus
help us make sense of the world. It does not, however,
follow that Statistical Independence is also a useful assumption in the quantum realm. And vice versa, assuming that Statistical Independence is
violated in quantum mechanics does certainly not imply that it must also be violated for mice and cancer cells. In fact, it is extremely 
implausible that a quantum effect would persist for macroscopic objects because we know empirically that 
this is not the case for any other quantum-typical behavior that we have ever seen. 

That it took 43 years for
someone to raise this blindingly obvious point goes to show what happens if you leave philosophers to discuss physics. 
Luckily, recently an explicit example was put forward \cite{Sudarsky} that shows how violations of Statistical Independence in 
a superdeterministic model for quantum mechanics are effectively erased along with decoherence. I sincerely hope  (alas, do not actually believe) 
that this will put to rest the claim that superdeterminism
is non-scientific.

Part of the confusion underlying this argument seems to be the belief that giving up Statistical Independence
is all there is to superdeterminism. But a violation of Statistical Independence is merely a property of certain models
whose ability to explain data needs to be examined on a case-by-case basis. Superdeterministic models do not
explain any arbitrary set of data thrown at them; they make specific predictions, to begin with those of quantum mechanics. 
The ultimate reason, however, to not assume Statistical Independence for
quantum objects is that this way we might be able to explain more data than we can with quantum mechanics. 
It is a deterministic approach that in principle predicts individual measurement outcomes, not just averages thereof. If someone came up with a
better explanation for cancer incidence in mice than Statistical Independence and a causal correlation between
tobacco exposure and cancer, then we should certainly take that seriously too.

Needless to say, the claim that superdeterminism would kill science has always been silly because most scientists care heartily little what is being discussed in the foundations
of quantum mechanics. And that's for good reasons: Quantum effects are entirely negligible for the description of macroscopic
objects.

\section{The relevance of future input for superdeterminism}

It may seem curious that the dynamical law whose properties we sketched in the previous sections
depends on the detector settings and yet has to be local. It is easy to see how one can
come up with a dynamical law that depends on the detector settings by postulating a
non-local interaction between the prepared state and the detector. However, doing so would
return us to pre-relativistic days and bring more
problems than it solves. We want a theory that is local in the
Einsteinian sense, ie one that does not have a ``spooky action at a distance''. But
the state at preparation has not actually been in contact with the detector, so how can it be
that its evolution depends on the settings?

As previously stressed, the superdeterministic 
evolution law depends on the measurement settings {\emph{at the time of
detection}}. There is no non-local interaction here. The prepared state need not interact
with the detector before measurement, it's merely that the setting appears in the evolution
law. For a simple example of such an evolution law, please refer to the accompanying
paper \cite{toymodel}. 

This behavior is sometimes referred to as ``retrocausal'' rather than superdeterministic, but I have
refused and will continue to refuse using this term because the idea of a cause propagating back
in time is  meaningless. Physicists
commonly define cause and effect by the causal structure of the underlying manifold. If
two events {\bf A} and {\bf B} are correlated, then {\bf A} is the ``cause'' of {\bf B},
if {\bf B} is in the forward lightcone of {\bf A}. If you think you have a situation where
{\bf B} ``retrocauses'' {\bf A}, then this merely means {\bf A} causes {\bf B}. 

The reason physicists use this space-time notion of causality is that it's the only notion of causality we know of
that fundamentally makes sense. The laws of physics, to the
extent that we have discovered them so far, do not have a direction of time. To best current
knowledge, then, the apparent
asymmetry of time is an emergent phenomenon that has its origin in the Past Hypothesis and
entropy increase. To
deal with this, we take a Riemannian manifold, assume it has a well-defined causal structure, and pick one 
direction as ``forward'' so that it agrees with the observed entropy increase.

Indeed, those who speak of causes going back in time are using a different notion of ``cause'', namely that used
in the analysis of causal discovery algorithms \cite{Pearl,Spirtes}. This notion of causality is operational and centers
on the usage of a model rather than on the model's mathematical basis. In this approach, causality is defined 
by studying the consequences of changes made by intervening agents.  This is best illustrated by an example: Why
do we say overeating is a cause of weight-gain? According to the intervention-based definition of causality
it's because if you stop binging on muffins you'll lose weight, whereas chopping
off your leg may reduce your weight but do nothing to stop you from munching down those muffins. 

Sounds reasonable enough. What's the problem with this? Well, the problem is that
``agents'' and their ``interventions'' are macroscopic terms that do not appear in quantum mechanics. 
Using them in quantum mechanics is equally bad as the measurement postulate, if not worse because the measurement postulate
at least does no harm. 

I seem to know a lot of people who are sold on this interventionist notion of causality, so let me add that I am
sure it is good for something, but whatever that something is, analyzing models in the foundations
of physics isn't it.\footnote{Tim Palmer wants to write a paper about this but he doesn't yet know it.} Case in
point: The authors of \ref{finetuning} define superdeterminism as a model in which ``the settings are not free but are
causally influenced by other variables'', using the interventionist notion of ``causal'' but then define retrocausation as
``the possibility of causal influences that act in a direction contrary to the standard arrow
of time'', ie explicitly using the space-time notion of causality. They thus leave out the option that the model may
be space-time causal but interventionist retrocausal, which happens to be the case discussed here. 

To be fair, this doesn't really matter for the argument of \ref{finetuning} which could well have included the
case discussed here. I am just picking on this to highlight that it's a bad idea to mix up two different
notions of causality. Before I explain what the real problem is with the argument of \ref{finetuning}, 
first let me say some more about future-input dependence. 

Future-input dependence is a term coined in \cite{Wharton} to describe retrocausality when using the interventionist notion
of causality. As emphasized above, this does not mean future-input dependence is retrocausal in the space-time way, so to avoid confusion,
I will simply refer to it as future-input dependence. Future-input dependence means that a model draws on 
input that will become available only in the future. 

Please keep in mind that the interventionist way to think about
causality is all about what you, as a user, can do with a model, not about the physics which the model
describes. So, saying that some input will only become available in the future does not mean it didn't previously
exist. In a deterministic model, if some information is present on one space-like hypersurface, it is present on all of
them. Thing is though, this information may not be available in any practically useful form before a certain point in time. 

This is exactly the situation with superdeterminism that we discussed in Section \ref{knots}. The input that goes into the superdeterministic
model is the measurement setting at the time of measurement. At the time of preparation, you
do not have this input, so it's fair to call it ``future input''. In principle, the information is there,
somewhere, on the preparation hypersurface, but for all practical purposes you have no way of figuring it out.
The result is that you have a model which makes contingent predictions depending on a parameter
you will only learn about in the future. You may not know at preparation what later
will be measured, but you can very well say: {\sl If}
the measurement setting is this, {\sl then} the prediction is that. 

You may wonder now why not take one of the good, old Hamiltonian evolutions, rather than one that
has future input in the dynamical law?
Because if you do this, the model isn't useful. You would have to precisely know the details
of the initial state to be able to calculate anything with it. Future-input dependence, hence, is the
reason why superdeterminism is not a conspiracy theory. It demonstrates that there is a simple
way to write down the dynamical law that does not require much information.\footnote{It
follows from this that a Hamiltonian evolution can at best be an effective limit of
superdeterminism, but not the full theory. We will come back to this point in Section \ref{outlook}.}

We can interpret future-input dependence entirely without the interventionist notion of causality
by rephrasing it in terms of a distance measures in configuration space. This point has been stressed in 
\cite{Hossenfelder:2019shy} and also appears in \cite{Hooft:2014kka}, if in a somewhat different form.
The connection is that you can define a ``small'' change of a state on a given hypersurface
as one that does not alter the detector settings on the hypersurface of measurement. Instead
of thinking of this as a distance-measure, you could also think of it as a probability-measure: Just
replace ``small'' with ``likely''. This gives you a well-defined and -- importantly -- simple way
of identifying initial states. Defining the measure in terms of the detector settings at a future time 
proves that you do not need to specify a lot of details to explain data with your model.

This is the reason why most finetuning
arguments against superdeterminism fail. They implicitly postulate that 
certain changes (say, in the digits of random generator) are ``small'' without ever
defining what this means. In the next two sections we will see explicitly how this mistake keeps creeping in.

\section{Superdeterminism may or may not allow superluminal signaling}
\label{suplum}

\begin{figure}[ht]
\vspace*{0.3cm}
  \begin{center}
    \includegraphics[width=0.88\textwidth]{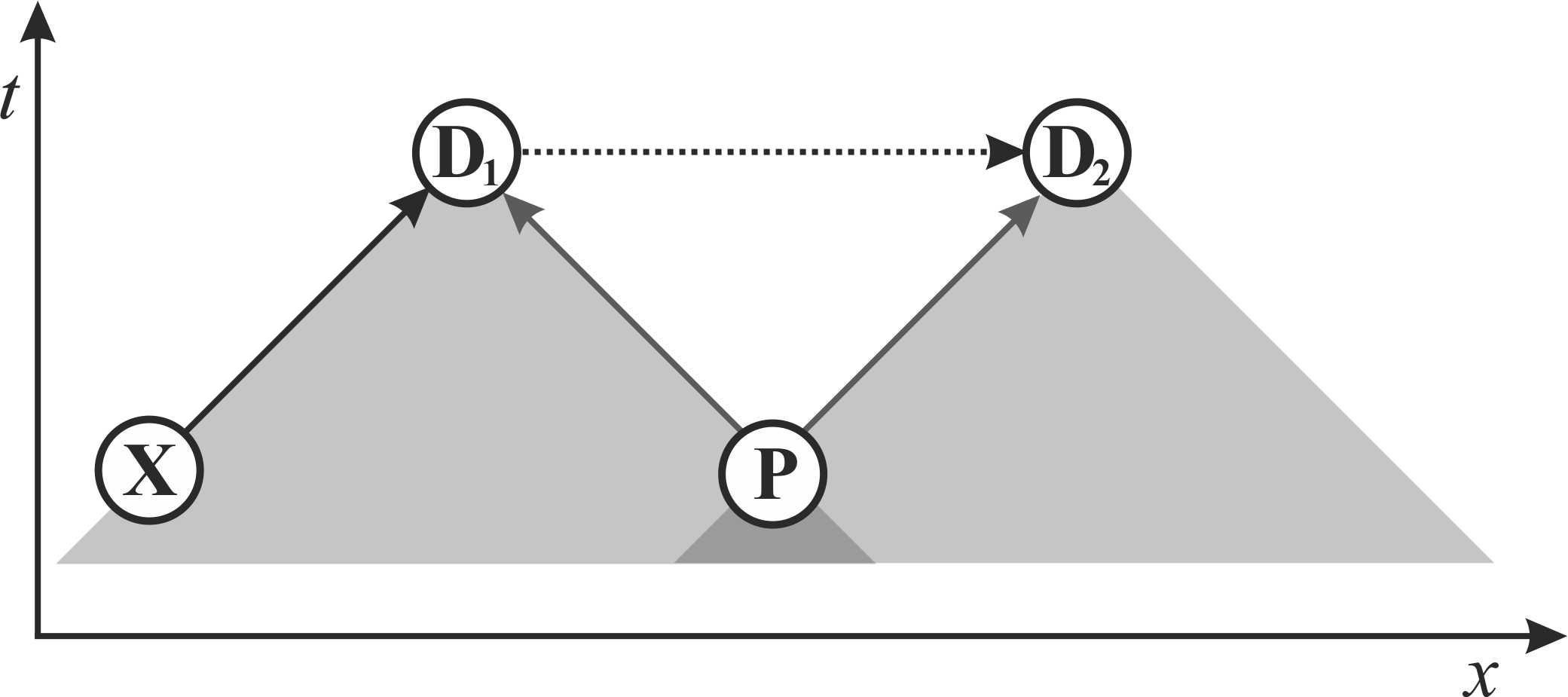}
  \end{center}
  \caption{Illustration for superluminal signaling. \label{fig1} }
\end{figure}

Superdeterminism generically allows superluminial signaling, where I am using the word ``generic'' in the mathematical sense, meaning a property that holds for typical examples, unless further details are specified. 

The relation between superdeterminism and superluminal signaling is most easily explained by the diagram in Figure \ref{fig1}. In a superdeterministic model, the outcome of the measurement prepared at {\bf P} and sent to the detector(s) depends on the detector settings at the time of measurement. The detector, however, can consist of multiple parts distributed over a space-like hypersurface, for example in points ${\bf D}_1$ and ${\bf D}_2$. The past lightcones of the two detectors have regions which do not overlap. If information from one of the non-overlapping parts, ${\bf X}$, is used for the detector setting ${\bf D}_1$, then this information can, in a superdeterministic model, in principle influence the outcome on the other detector ${\bf D}_2$. 

So, if the only thing you know about a model is that it is superdeterministic, then superluminal signaling cannot be excluded. But one cannot tell whether information can be submitted superluminally in a given model unless one specifies details of the model. The question whether and how much superluminal signaling is possible can therefore not be answered in general for all superdeterministic models. There are however a few general statements that can be made.

First, and most important, superluminal signaling is problematic only when it is in conflict with observation or when it leads to internal inconsistencies. Superluminal signaling is a priori compatible with general covariance, and it does not result in time-like closed loops as long as one has an arrow of time (defined, eg, by entropy increase), which (for better or worse) our universe does have. It is therefore not a show-stopper, but merely an eyebrow-raiser. 

Second, we know already that not all superdeterministic models allow for superluminal signaling because the hidden variables may be uncomputable from observational evidence and thus cannot be exploited for messaging. This is the option advocated eg in \cite{Palmer:2008jh}. In this case, the measurement outcomes remain for all practical purposes random, though they are determined in principle. (This does not mean the model has no empirical relevance, because the underlying structure may lead to other deviations from quantum mechanics, for example by predicting a maximal number of particles that can be entangled \cite{Palmer2018}.) 

Third, even if the hidden variables are computable in principle, it may be impossible to transfer information with knowledge about them in practice because the outcome of the measurement is extremely sensitive to the detailed experimental setup. 

An example for how the latter issue tends to be neglected is the scenario of  \cite{Bendersky}. In this scenario, we have Alice and Bob, each with a detector. They are making measurements on entangled pairs of particles from a source somewhere between them. The authors of  \cite{Bendersky} then imagine a ``learning round'' in which Alice and Bob each do their measurements and share the results, after which these results are being exploited in the ``signaling round''.

The key point of the argument in \cite{Bendersky} is a dedicated learning algorithm (which assumes a specific type of computability of the hidden variables), but the details don't matter here. What matters here is merely that Alice and Bob can collect information about the underlying deterministic law from subluminally exchanging some amount of measurement outcomes. And once they have information about just how the time-evolution is determined (even if this information is incomplete), it seems they can figure out what to do on one end -- say, Bob's -- to get a specific, not a random, result on Alice's end. 

The trouble is that the learning round does not necessarily allow anyone to  superluminally signal anything. Keep in mind that the theory is deterministic. If Alice and Bob have exchanged enough information, they will eventually just know enough about the initial state of the entire system to predict the future evolution (or at least certain aspects thereof). But at this point they already have all the information there is to have and nothing is left to signal. 

To be able to actually send information, Bob would have to {\emph{change}} the setup on his side, so that Alice's outcome depends on the new information he injects into the system. However, by doing so he screws up the knowledge he has gained during the learning round. The only information he can send is thus the information he put into his setup before the learning round, and so it would not be transferred superluminally.

One may object that we do not know whether Bob's attempt to inject information into the system will actually wipe out his ability to message to Alice. Maybe there is a way to avoid this if he is careful enough. Yes, maybe. Again, this is not a question which can be analyzed without knowing further details of the superdeterministic model. However, that small changes can significantly affect the future time evolution is a property we expect of a superdeterministic model exactly because the deterministic nature of whatever underlies quantum mechanics has been so difficult to discover. This difficulty means that the measurement outcome almost certainly depends very sensitively on the exact setup of the system. It adds to this that we know the theory has to be non-linear and the time-evolution is quite possibly chaotic. 

This leads one to suspect it is not the absence of superluminal signaling that requires fine-tuning but, to the very contrary, it is {\emph{succeeding}} in superluminal signaling that requires fine-tuning. This property of superdeterminism is implicit in the very fact that Born's rule does not depend on what else is going on in the rest of the universe. Born's rule is what one gets if one does not know the details. It follows that the details are what determines exactly which prepared state leads to which measurement outcome. So, only if one has control over the details -- ie, if one finetunes -- then can one predict the outcome, and, at least in principle, message with it. 

The bottom line is that the learning algorithm of \cite{Bendersky} does not tell us anything about the possibility to signal superluminally. Merely being able to predict measurement outcomes does not equal signaling.

\section{Superdeterminism is not necessarily finetuned}
\label{finetuning}

A ``finetuned'' theory or a ``conspiracy theory'' is a theory which lacks explanatory power. 
A lack of explanatory power means that the theory requires more information as input than just collecting the data. 
This notion of finetuning makes sense because a theory that is finetuned in this way is unscientific. We should
prefer just collecting the data over such a theory. 

There are several other notions of ``finetuning'' that have been used in the foundations of 
physics. These other notions of finetuning, however, are meta-physical, meaning they have no empirical justification \cite{Hossenfelder:2018ikr}. 
One can certainly define these other notions of finetuning, but there is no reason to think that a theory finetuned in such
a way is problematic. Unfortunately, those who use other notions of finetuning are typically not aware that
their definitions do not support their conclusion (that the theory is unscientific). 
The origin of this confusion is always
the same: They implicitly assume knowledge of a probability distribution that is unknown or, in some
cases, unknowable.

Consider this common example of finetuning: A pen balanced on its tip. Should we encounter such
a situation, we would suspect a hidden mechanism because otherwise the balance seems ``unnatural'',
a ``conspiracy'', or ``finetuned''. The reason is that we know the system is unstable under perturbations that
(here comes the relevant point) are likely to occur (during the time of observation) given the initial
conditions we are likely to encounter. This means we have an
empirically supported expectation for the state the system is likely in, and the perturbations it is
likely to be exposed to, and these experiences have themselves been obtained by drawing on theories which
have explanatory power. 

For this reason, seeing
a precariously balanced pen leads us to discard the possibility that it just happens
to have had the right initial condition as possible but implausible. It is much
more likely the pen is held in place by something. Any mechanism that did that would
remove the need to finetune and thus have much higher explanatory power. 

The balanced pen is an example for a good, scientific finetuning argument. As example for an unscientific finetuning
argument take the supposed issue that certain parameters of the standard model of
particle physics are finetuned. In this case, we have no way to quantify the probability of having a
different parameter in this model because that would describe a universe we do not inhabit. Neither
can these parameters be perturbed in the real world.\footnote{One can certainly turn the parameters
into variables, but this would just be a different model.} The consequence is that this notion of
finetuning used to criticize the standard model has no empirical basis. And unsurprisingly so: The standard model is
in some sense the scientifically most successful theory
that humans have ever conceived of. It explains Terabytes of data with merely some dozen
parameters. It would be hugely absurd to complain it lacks explanatory power.\footnote{In 
some areas of particle physics and cosmology this already unscientific notion of finetuning is aggravated by furthermore
applying it to constants are unobservable to begin with, making it even more detached from empirical relevance. Typically one
considers in such arguments constants that are terms whose sum is an observable. I.e., if $x$ and $y$ are numbers that appear in
 intermediate steps of a calculation, but only $x-y$ is observable, then worrying about $x$ and $y$ is not
scientifically justified because these constants have no physical meaning in and by themselves. For more
details on problems with finetuning arguments in cosmology and particle physics, please see \cite{Hossenfelder:2018ikr}.}

This is not to say that finetuning arguments are always unscientific. Finetuning arguments can be
used when one has a statistical distribution over an allowed set of parameters and a dynamical law
acting on this distribution. If the initial distribution requires us to use detailed information
to get the observed outcome then the theory becomes scientifically 
useless because that amounts to assuming what one observes.\footnote{This does not mean the theory is wrong. It just means it's not scientific.}

One can shortcut the philosophical considerations about finetuning and just ask: What can I do with the model? If the model allows
you to correctly calculate measurement outcomes, it has explanatory power, hence is not finetuned. If
you are yourself unsure how well a model explains data, it helps to look at what researchers in practice 
spend their time on. You can tell, e.g., that supersymmetry does not
remove any supposed finetuning issues with the standard model because no one who evaluates standard model cross-sections has
use for adding parameters that don't do anything to ease calculations or fit data. 
 
A common side-effect of using a scientifically meaningless notion of finetuning is that models with lesser explanatory power will appear superior. This is because, once one falsely flags a model as finetuned, scientists will try to ``fix'' it, typically by adding assumptions that have the effect of making the model more complicated, hence lowering its explanatory power. 

A good example for this is the cosmological constant. The cosmological constant, for all we currently know, is a constant of nature. I find it hard to think of an assumption for a scientific theory that has higher explanatory power than a regularity expressed by a constant of nature. In the context of the concordance model, it explains all available data.\footnote{Granted, there may be some tension with the recent data, but this isn't the point here.} However, if you erroneously think there's something wrong with a cosmological constant because you implicitly assumed an empirically unknowable probability distribution over the constant's values, you may try to ``improve'' on the concordance model by inventing a complicated ``explanation'' for why the
constant has the value it has. I am scare-quoting these terms because adding assumptions that do nothing to improve the fit to data do not increase, but decrease, explanatory power. Again, one can see this just by noting that no one who actually makes calculations with the concordance model starts with some superfluous story about, say, multiverses, just to in the end, as always, use the best-fit value for the cosmological constant.

With that ahead, let us return to superdeterminism and see how it fares on the accusation of
being finetuned that has been raised e.g. in \cite{finetuning,Valentini1,Valentini2}. These finetuning
arguments, it must be noted, differ from the earlier mentioned ones. They do not claim that superdeterminism
is finetuned per se, but that superdeterminism is finetuned {\emph{if}} one wants to exclude the
possibility of superluminal signaling.

The authors of \cite{finetuning} avoid the problem with \cite{Bendersky}'s argument (see Section \ref{suplum}) by not telling the reader how the superluminal signaling can be achieved to begin with. They instead just claim that superdeterminism is problematic using a particular notion of finetuning based on the ``faithfulness'' of a causal model (actual quote, not a scare-quote). The causal model is ``faithful'' or ``not finetuned'' if ``its conditional independences continue to hold for any variation of the causal-statistical parameters''. This definition leaves aside whether the parameters of the model can be varied to begin with, or if, how likely such variations are. I.e., the definition suffers from the very problem we discussed above. 

That the notion of finetuning in \cite{finetuning} is scientifically meaningless is apparent by the authors' elaboration on Bohmian mechanics. Bohmian mechanics with the equilibrium hypothesis is finetuned, according to their definition, because one could vary the equilibrium hypothesis, which allows deviations from Born's rule and thus potentially also allows superluminal signaling. But given that Bohmian mechanics with the equilibrium hypothesis is equivalent to the Copenhagen Interpretation with Born's rule, this means in \cite{finetuning}'s terms, quantum mechanics itself is finetuned. 

Now, quantum mechanics is arguably a theory with high explanatory power, so what is going wrong with their argument? What goes wrong is that they do not quantify the probability for conditional independences to not hold.\footnote{In fact, the biggest part of the paper \cite{finetuning} is dedicated to explaining why the causal discovery algorithms that they use are not suitable to analyze entanglement because, in their own words, ``Independences simply do not provide enough information. One needs a causal discovery algorithm that looks at the {\emph{strength of correlations}} to reproduce Bell's conclusions.'' (Emphasis original.)} They do not do this because they cannot: Doing so would require specifying a probability distribution for varying the model's parameters. We do not have such a distribution because we have no data for superdeterminism except that which supports Born's rule. And Born's rule itself is a good example for an assumption that removes finetuning. The reason Born's rule is not finetuned is that it is simple and it explains a lot of data. This is correct in quantum mechanics, and is still correct in superdeterminism, because it is correct, period. We know this empirically.

I should add here, that the claim in \cite{finetuning} is, more precisely, that superdeterminism (and retrocausality) is {\emph{more}} finetuned, according to their definition, 
than quantum mechanics (and Bohmian mechanics, and many worlds, and any other logically equivalent formulation of quantum mechanics). But since we have just seen that their definition tells us nothing about explanatory power, there isn't much to learn from this. It's like they first demonstrate a detector doesn't measure altitude but then use the same detector to claim the Rocky Mountains are higher than Mt.\ Everest. 

The reader who doubts that a superdeterministic theory can reproduce the predictions of quantum mechanics without finetuning is welcome to have a look at the toy model put forward in \cite{toymodel}. It does away with the collapse postulate by positing a violation of Statistical Independence. Finetuning is not required because all it takes to get Born's rule is the detector setting, not the details of the hidden variables.
In the same limit where the toy model reproduces quantum mechanics, it also does not allow for superluminal signaling. 

It was also previously demonstrated in \cite{rfinetuning} that finetuning in a superdeterministic model can alternatively be avoided by
drawing on symmetry constraints, and in \cite{Hossenfelder:2020gdb} it was further shown that Born's rule can generally be derived from a symmetry requirement
which provides a simple (hence, not finetuned) explanation for why quantum mechanics is robust. This type of argument should be familiar to particle physicists. 

It was mentioned in \cite{finetuning} that an evolution law with attractor behavior can remove finetuning. This is sometimes called ``dynamical relaxation'' in the scientific literature, when applied to certain initial conditions. The idea here is that a dynamical law with an attractor can make a model less sensitive to the choice of initial conditions. But note that this is the type of superfluous assumption that we discussed above. Born's rule, or the equilibrium hypothesis for pilot waves, are simple assumptions that suffice to explain observations -- so far. Replacing them with a dynamical law that does not improve the fit to data decreases the explanatory power of the model.

However, as emphasized earlier, one does not expect the limit in which superdeterminism recovers Born's rule to be exact. Deviations from it should be experimentally observable, eventually. Therefore, one day the additonal assumptions may become justified because they improve the capability of the model to fit data. But presently, we have no information on which to base the assumption that deviations are somehow large, or likely, or even that they can be exploited in practice. 

The argument put forward in \cite{Valentini1,Valentini2} suffers from exactly the same problem as that in \cite{finetuning}. The authors of \cite{Valentini1,Valentini2} postulate that certain variations of parameters in superdeterministic models must be possible and are likely. They fail to notice that any statement of this type would require knowledge of both the model and the probability distribution.

To summarize this section: A model is finetuned if it lacks explanatory power. To evaluate whether a model is finetuned, one therefore needs to look at a model.  

\section{Outlook}
\label{outlook}

The two biggest problem with superdeterminism at the moment are (a) the lack of a generally applicable fundamental theory and (b) the lack of experiment. 

The lack of experiment could be remedied with existing technological capacities even in absence of a fully-fledged theory if the quantum foundations community could overcome their obsession with Bell-type tests. Bell-type tests will never allow us to tell superdeterminism apart from quantum mechanics. Instead, we need to look for deviations from Born's rule in small and cold systems \cite{Hossenfelder:2019shy,Hossenfelder:2011ct}. 

For what the theory development is concerned, what is required is a mathematical formalism that will give rise to a non-linear evolution law of the type discussed above, where the locations of the attractors depend on the detector settings. The difficulty is that the detector settings themselves are degrees of freedom in the model. 

Decoherence theory \cite{Zeh1,Zurek1,Zurek:2003zz} tells us how to identify the relevant pointer states of the detector, so this is not the difficulty. The difficulty is that these have to effectively appear in the dynamical law and play the role of attractors. As mentioned earlier, it is hard to see how one can formulate such a theory with the usual quantum mechanical Hamiltonian formalism because it is a first-order differential equation. Given an initial state, there's no place to account for future input. This agrees with what we noted above that it is not quantum mechanics which we want to make non-linear. We are looking for a theory from which quantum mechanics derives on the average.

The toy model put forward in \cite{toymodel} should be understood as a effective limit of a superdeterministic theory. In this effective limit, the detector settings are hard-coded into a modified Hamiltonian evolution. The theory we are looking for would explain how one obtains such an effective limit. 

I believe this theoretical challenge can be tackled with existing mathematical methods. What prevents us from making progress here is simply that we haven't made enough effort. Because superdeterminism has historically been stigmatized as ``unscientific'', pretty much no work went into formulating a useful theory for it. So, at present all we have are toy models. I sincerely hope that these notes will serve to encourage some readers to explore the benefits of giving up Statistical Independence and make headway on developing a theory for superdeterminism.


\section*{Acknowledgements}

I thank Sandro Donadi and Tim Palmer for helpful discussions. This work was supported by the Fetzer Franklin Fund.

\end{document}